# Fractional Debye-Stokes-Einstein behaviour in ultraviscous nanocolloid: glycerol and silver nanoparticles

[1,2]Szymon Starzonek, [1,2]Sylwester J. Rzoska(*), [2]A. Drozd-Rzoska,

[1]Sebastian Pawlus, [1]Ewelina Biała,

[3]Julio Cesar Martinez-Garcia and [4]Ludmila Kistersky.

[1]Silesian Intercollegiate Center for Education and Interdisciplinary Research &Institute of Physics, University of Silesia, ul. 75 Pułku Piechoty 1A, 41-500 Chorzów, Poland

[2]Institute of High Pressure Physics, Polish Academy of Sciences, ul. Sokołowska 27/39, Warsaw 01-142, Poland.

[3]University of Berne, Freiestrasse 3, Berne CH-3012, Switzerland.

[4]V. Bakul Institute for Superhard Materials of the National Academy of Superhard materials NASU, Avtozavodskaya Str.2, 04074 Kiev, Ukraine

(*) Corresponding author: sylwester.rzoska@gmail.com

**Abstract**

One of hallmark features of glass forming ultraviscous liquids is the decoupling between translational and orientational dynamics. This report presents studies of this phenomenon in glycerol, a canonical molecular glass former, heading for the impact of two exogenic factors: high pressures up to extreme 1.5 GPa and silver (Ag) nanoparticles (NP). The analysis is focused on the fractional Debye-Stokes-Einstein (FDSE) relation $\sigma(T,P)(\tau(T,P))^S = const$, linking DC electric conductivity ($\sigma$) and primary (alpha, structural) relaxation time ($\tau_\alpha$). In glycerol and its nanocolloid (glycerol + Ag NP) under atmospheric pressure only the negligible decoupling ($S \sim 1$) was detected. However, in the compressed nanocolloid a well-defined transformation (at $P$ = 1.2 GPa) from $S \sim 1$ to the very strongly decoupled dynamics ($S \sim 0.5$) occurred. For comparison, in pressurized 'pure' glycerol the stretched shift from $S \sim 1$ to $S \sim 0.7$ took place. This report presents also the general discussion of FDSE behavior in ultraviscous liquids, including the new link between FDSE exponent, fragility and the apparent activation enthalpy and volume.

**Introduction**

**I. Conceptual background**

Glass transition physics has remained a challenge for condensed and soft matter physics since decades.[1-3] The most intriguing feature is the set of strong previtreous effects for dynamic properties, with similar patterns for qualitatively different glass forming systems.[2] The key representative of such behavior is the super-Arrhenius (SA) evolution of various dynamic properties on approaching the glass temperature $T_g$:[2,3]

$$x(T) = x_0 \exp\left(\frac{\Delta E_a(T)}{RT}\right), \qquad T > T_g \qquad (1)$$

where $x(T)$ stands for the primary (*structural, alpha*) relaxation time ($\tau_\alpha$), viscosity ($\eta$), diffusion ($D$) or reciprocal of DC electric conductivity ($1/\sigma$). $\Delta E_a(T)$ denotes the apparent activation energy, $T_g$ is for the glass temperature and $R$ is the gas constant. The basic Arrhenius equation can be restored for $\Delta E_a(T) = \Delta E_a = const$.

The 'universal' metric of the SA behavior is called 'fragility' and defined as follows:[2,4]

$$m = m_{P=const} = \left[\frac{d\log_{10} x(T)}{d(T/T_g)}\right]_{T=T_g} \qquad (2)$$

It ranges from $m \approx 16$ for the 'clear' Arrhenius behavior to $m \sim 200$ for the strongly SA dynamics. The fragility is considered as one of the most important parameters of the glass transition physics: the metric linking microscopically distinct systems, including low molecular weight liquids (LMW), polymers (P), colloids…..[2,5] Notwithstanding, its fundamental meaning is still well characterized by the title of ref.[6]: '*The fragility and other properties of glass-forming liquids: Two decades of puzzling correlations*''. Only recently, a clear link to basic process energies has been derived:[5] $m = C(\Delta H_a / \Delta E_a)_{T=T_g}$, where $C = \log \tau_\alpha(T_g) - \log \tau_0$, $\Delta H_a = d\ln \tau_\alpha(T_g)/d(1/T_g)$ is the activation enthalpy, $\tau_\alpha(T_g) = 100\,s$ and $\tau_0 = 10^{-11}s - 10^{-16}s$ is the prefactor in the SA eq. (1). The SA behavior is assisted by the stretched exponential (SE) time-decay of physical properties $I(t) \propto \exp[-(t/\tau_\alpha)^\beta]$, with the SE exponent $0 < \beta \leq 1$ or equivalently the non-Debye distribution of relaxation times in the frequency domain.[2]

It is particularly notable that the evolution of translation and orientation related dynamic properties ($x(T)$) in the ultraviscous domain near $T_g$ is decoupled, what manifests via fractional Stokes-Einstein (FSE) and Debye-Stokes-Einstein (FDSE) relations:[2,7-10]

$$D/T = A\eta^{-1+\varpi} \quad \text{and} \quad D/T = A'\tau_\alpha^{-1+\varpi} \qquad \text{with exponent} \quad \varsigma = 1-\varpi \qquad (3)$$

$$\sigma\tau_\alpha^{1-\mu} = const \qquad \text{with exponent} \quad S = 1-\mu \qquad (4)$$

where $A$ and $A'$ are constants

The experimental evidence indicates that generally FDSE or FSE behavior with non-zero fractional exponents ($\varpi, \mu \neq 0$) takes place in the ultraviscous/ultraslowing dynamical domain for $T_B(\tau_\alpha \sim 10^{-7}s, \eta \sim 10^3 Poise) < T(\tau, \eta) < T_g(\tau_\alpha \sim 10^2 s, \eta \sim 10^{13} Poise)$. For $T > T_B$ the crossover to DSE or SE behavior occurs ($\varpi, \mu = 0$ or $\varsigma, S = 1$).[2,8-40] The temperature $T_B$ is related to the crossover from the high temperature (ergodic) to the low-temperature (non-ergodic) dynamic domain.[2,11] The latter is also associated with the appearance of multimolecular dynamic heterogeneities (DH) or alternatively cooperatively rearranging regions (CRR) near the glass transition, with vastly different relaxation times and viscosity.[2,16-40] They are considered as the most probable reason for universal patterns in the ultraviscous/ultraslowing (low temperature) domain in the immediate vicinity of the glass transition.[2] Studies of FDSE or FSE behavior are recognized as one of key tools for getting insight into still mysterious dynamic heterogeneities.[2,8,15] Notwithstanding, the knowledge regarding the fundamental background of FDSE/FSE behavior is still heuristic, despite the growing up number of experimental and theoretical research reports.[2,7-40]

All these suggests the significance of FDSE studies in ultraviscous liquids atemporal "*research status quo*". This is the target of the given report.

First, the resume of FDSE reference results, particularly focusing on eq. (4), is presented. This topic is concluded by the novel link between the FDSE exponent and basic characteristics of the SA behavior, namely: the fragility, the activation enthalpy and the activation volume. Second, results related to the impact of exogenic factors on dynamics of glycerol, one of canonical glass forming liquids, are presented. They are: (i) high pressures up to challenging *P*=1.5 GPa and (ii) the addition of silver nanoparticles (Ag NP), forming a nanonocolloid/nanocomposite/nanofluid to the ultraviscous glycerol. The impact of nanoparticles lead to the crossover to the strongly decoupled region in the immediate vicinity of the glass transition (i.e. within the ultraviscous domain), a phenomenon which has been not reported before.

**II. The translational-orientational decoupling**

For coupling between translational and orientational processes in 'classical' liquids one can expect the validity of Debye, Stokes and Einstein relations:[2,14,41]

$$\frac{D_{tr}}{T} = \frac{k_B}{6\pi r_{ion}}\eta^{-1} \quad (5), \qquad \tau_\alpha = \frac{v\eta}{k_B T} \quad (6), \qquad \frac{D_{rot}}{T} = \frac{k_B}{8\pi r_{dip}^3}\eta^{-1} \qquad (7)$$

where $D_{tr}$ and $D_{rot}$ denote translational and rotational diffusivities, $v = A'V$, $r$ the radius of diffusing molecule and $V$ is for the molecular volume.

It is notable that for the Debye-Stokes (DS) eq. (6) $\tau_\alpha(T) \propto \eta(T)/T$.[15,41] However, the alternative approach via the Maxwell relation[15] yields $\tau = G_\infty \eta$. Consequently, assuming that in the ultraviscous domain the instantaneous shear modulus $G_\infty = const$ one obtains $\tau(T) \propto \eta(T)$.[15] It is worth recalling that in the Maxwell relation $\tau$ denotes the stress relaxation time and there are no clear experimental evidence that the structural ($\tau_\alpha$) and stress relaxation time ($\tau$) are interchangeable.[2,15] Linking above dependences with the Nernst-Einstein (NE) relation $D_{tr} = k_B T\sigma/nq^2$,[14] where $n$ is the number of electric charges/carriers, $\sigma$ denotes the DC electric conductivity and $q$ is the electric charge, one obtains:

$$\sigma\tau_\alpha = \frac{nq^2 Cv}{k_B T}, \text{ i.e. } T\sigma\tau_\alpha = const \quad (8) \quad \text{or} \quad \sigma\tau_\alpha = \frac{CG_\infty ne^2}{a}, \text{ i.e. } \sigma\tau_\alpha \approx const \quad (9)$$

where eq. (8) recalls DS eq. (5) and eq. (9) is based on the Maxwell equation, as discussed above.

In low molecular weight liquids the DC conductivity arises from residual ionic dopants: salts or other ionic species that inevitably can get into samples during the synthesis.[12] For broad band dielectric (BDS) spectra such behavior always dominates at lower frequencies, often beginning just below the kHz domain. In ionic or highly conductive liquids this can be the governing factor also for the multi MHz region. It is notable that taking into account the Nernst-Smoluchowski (NS: $D_{tr} = \lambda/2\tau_h$ )[14,41,42], and the NE equations one obtains the relation linking DC conductivity and the hoping time of ions, responsible for the DC conductivity, namely:

$$\sigma = \frac{nq^2}{kT} \frac{\lambda}{2\tau_h} \quad (10)$$

where $\tau_h$ is the hoping length of diffusing species, $n$ is free ions concentration and $q$ is ion charge.

This relation makes it possible to present eqs. (3) - (7) as the result of the comparison between two time scales associated with the orientation of molecules (~ primary, alpha, structural relaxation) and the translation related ions hopping time. The entrance into the ultraviscous domain converts eq. (10), NS and NE relations into their fractional forms. Consequently, FSE and FDSE eqs. (3) and (4) can be presented as the results of the comparison of two mentioned time-scales:[43 and refs. therein]

$$\frac{\tau_\alpha}{\tau_h} \propto (\tau_\alpha)^\mu \qquad \text{and} \qquad \frac{\tau_\alpha}{\tau_h} \propto (\tau_\alpha)^\varpi \qquad (11)$$

what suggest that $1/\tau_h \propto \sigma$ and exponents $\mu = \varpi$.

The last dependence resembles the one used in polymeric systems for the comparison of segmental ($\tau_S$) and chain ($\tau_C$) relaxation processes: $R = \tau_S/\tau_C \propto (\tau_S)^\varepsilon$. For low molecular weight liquids (LMW) the primary relaxation time $\tau_\alpha$ can be compared to $\tau_S$ and $\tau_C$ to the large time scale: in polymers it is estimated via $\tau_C \propto \langle \tau_S \rangle$. In ref.[43] the thermally activated barrier hoping model for the glass transition phenomenon was recalled to discuss deeper this issue. This model assumes the leading role of heterogeneities/(local domains) coordinating a group of molecules (LMW) or segments (P) in the ultraviscous/ultraslowing region. Fluctuating local density excesses results in a distribution of barrier heights, which gives rise to the decoupling of primary relaxation time and diffusion related processes as well as to the stretched exponential (non-Debye) relaxation. For polymers this 'heterogeneous' model yields: $\varepsilon = \Delta/(1+\Delta) = 1 + (1/a_c q)$, where $\Delta = a_C q' = a_C \sigma_E^2/2\langle F_B \rangle$, $\sigma_E^2$ is the energy barrier fluctuations variance, $\Delta E_a(T) = a_c F_B(T)$ in SA eq. (1), $a_c$ is a presumably temperature independent cooperation parameter, $F_B(T)$ is the hopping barrier energy and $\langle F_B(T) \rangle$ is for its mean value. The parameter $q' = 0.1 - 0.2$ is for the volume fraction of cooperative domains (heterogeneities). Basing on above dependences and the semi-empirical correlation for the fragility $m = 16 + 40.6(a_c)^{0.56}$ in polymers one can directly arrive to the relation which can be easily tested experimentally:[43]

$$\frac{1}{\varepsilon} = \frac{q'}{[(m-16)/40.6]^{0.56}} \qquad (12)$$

The compilation of experimental data for polymeric glass formers confirmed the smooth dependence of $\varepsilon$ vs. fragility $m$ predicted by the above relation. The important result of ref.[43]

was that the chain relaxation and fragility should weakly depend on the material as well as be insensitive to local heterogeneities due to the large-scale averaged nature of $\tau_C$. This behavior is in strong contrast to 'segmental' related dynamics ($\tau_S$), where there are notably SA behavior and fragility (large *m* values). All these can be associated with local cooperativeness ('heterogeneities'). The authors of ref.[43] suggested the same scenario for non-polymeric ultraviscous liquids, what leads to the equivalence: $\tau_C \rightarrow \tau_h (\propto 1/\sigma, 1/D_{tr})$ and $\tau_S \rightarrow \tau_\alpha$ in polymers and LMW. It is notable that the link of fractional decoupling exponents to dynamic heterogeneities is the output result of various glass transition models.[43] However, eq. (12) offers a unique possibility of experimental tests of such hypothesis. Notwithstanding, it is also associated with a notable arbitrariness, namely: (i) it includes the assumption that the prefactor in eq. (1) is universal ($\tau_0 = 10^{-14} s$ and then $C = 16$), (ii) the average energy $\langle F_B \rangle$ is poorly defined due to strong changes within the ultraviscous domain and for different glass formers.

In experimental studies on utraviscous low molecular weight liquids (LMW) particular attention attracted eq. (4) linking structural relaxation time and DC electric conductivity. So far, this is the only 'fractional coupling" relation which can be tested both as the function of temperature and pressure. Moreover, experimental values of $\sigma(T,P)$ and $\tau_\alpha(T,P)$ can be determined from the same scan of the imaginary part of dielectric permittivity $\varepsilon''(f)$, using the broad band dielectric spectroscopy (BDS). This fact essentially reduces biasing artifacts. There is a broad experimental evidence supporting the validity of FDSE eqs. (4), (8), (9) and showing that the exponent $0.75 < S < 0.9$.[2,10,16-26,37-41] Psurek et al.[20,23,24] indicated a possible pressure-temperature isomorphism for the FDSE behavior, namely:

$$\sigma(T,P)\tau_\alpha(T,P)^{1-\mu} = const, \quad 1-\mu = S \qquad (13)$$

It is notable that pressure studies focused on testing eq. (13) were carried out for relaxation times $10^{-6}s < \tau_\alpha < (10^{-1}s - 10^{-3}s)$ for near room temperatures and moderate pressures $P < 0.3 GPa$.[2,10,16-26,37-41] Such limitations resulted from still existing frequency restrictions in high pressure BDS studies.[13] Notwithstanding, the tested time-scale in pressure studies was well located within the ultraviscous and low temperature domain, adjusting to the glass transition at $(P_g, T_g)$.[2,11]

When discussing the FDSE behavior in glass forming ultraviscous liquids worth recalling is the challenging compilation of experimental data for 50 glass forming liquids focused on the normalized version of FDSE eq. (4).[8] For all liquids in the ultraviscous domain the same 'universal' FDSE exponent $\zeta \approx 0.85$, i.e. $\omega \approx 0.15$ was obtained.[8] This analysis included glycerol, which is the object of the given report. It is notable that linking eqs. (3), (4) and (8) one obtains:

$$\sigma \tau_\alpha^\zeta = \frac{CG_\infty^{-\zeta} nq^2}{a}, \quad \text{i.e. the fractional exponent } S = \zeta \qquad (14)$$

This result suggests the hypothetical equivalence of all discussed above FDSE power exponents. Although the vast majority of experimental evidences support the appearance of FDSE behavior in ultraviscous glass formers, or even more generally in highly viscous soft matter/complex liquids systems, results indicating a gradual decrease of FDSE exponents also exist. Worth recalling are also controversies related to the question whether the FDSE behavior is described via $T\sigma\tau_\alpha^S \approx const$ [2,15] or $\sigma\tau_\alpha^S \approx const$ [2,13-40]. The prevalence of the evidence supporting the latter dependence is most often explained via the statement that in the tested range of temperatures in ultraviscous liquids the change of temperature is small and negligible.[13, 16-18] In the opinion of the authors this 'general claiming' poorly coincides with the fact that the ultraviscous domain extents up to even $\Delta T \approx 100K$. Regarding this fundamental issue, the discussion related to eqs. (8) and (9) indicates that the dependence $\sigma\tau_\alpha^S \approx const$ is related to the 'elastic" Maxwell model with $G_\infty = const$, and such behavior seems to dominate in the ultravicous domain. One can expect that inherent features of the Debye model causes that the relation $T\sigma\tau_\alpha^S \approx const$ may be valid in the high temperature domain.

Generally the pressure counterpart of SA eq.(1) is given by:[2,13,42]

$$x(P) = x_0^P \exp\left(\frac{P\Delta V_a^x(P)}{RT}\right) \quad , \quad T = const. \tag{15}$$

where $\Delta V_a(P)$ is the apparent activation volume ('*free volume*').

It can be called super-Barus (SB), since the basic equation proposed by Barus[44] $x(P) = x_0^P \exp(cP)$ can be rewritten as $\Delta V_a^x/RT = c = const.$ The SA eq. (1) enables determining of the apparent activation enthalpy via $\Delta H_a^x(T)/R = d \ln x(T)/d(1/T)$.[45-47] Following refs.[2,45-47] the SB eq. (15) yields the apparent activation volume via $\Delta V_a^x(T)/R = d \ln x(T)/d(1/T)$. Then, basing on the FDSE eq. (13) one obtains:

$$\begin{cases} -\dfrac{d \ln \sigma(P)}{dP} + S\dfrac{d \ln \tau_\alpha(T)}{dP} = -\dfrac{\Delta V_a^\sigma(P)}{RT} + S\dfrac{\Delta V_a^\tau(P)}{RT} = 0 \Rightarrow S = \dfrac{\Delta V_a^\sigma}{\Delta V_a^\tau} \\ -\dfrac{d \ln \sigma(T)}{d(1/T)} + S\dfrac{d \ln \tau_\alpha(T)}{d(1/T)} = -\dfrac{\Delta H_a^\sigma(T)}{RT} + S\dfrac{\Delta H_a^\tau(T)}{RT} = 0 \Rightarrow S = \dfrac{\Delta H_a^\sigma}{\Delta H_z^\tau} \end{cases} \tag{16}$$

Consequently, for the given point in the (*P, T*) plane:

$$S = \frac{\Delta H_a^\sigma}{\Delta H_a^\tau} = \frac{\Delta V_a^\sigma}{\Delta V_a^\tau} \tag{17}$$

Direct implementations of the SA eq. (1) or SB eq. (15) for portraying experimental data are not possible, due to unknown forms of the apparent activation energy and volume. Consequently, *ersatz* dependences are used. The dominant is the Vogel-Fulcher-Tammann (VFT) relation[2,12] or its pressure related quasi-counterpart introduced in ref.[48]:

$$x(T) = x_0^T \exp\left(\frac{D_T T_0}{T - T_0}\right) \qquad (18) \qquad x(P) = x_0^P \exp\left(\frac{D_P P_0}{P_0 - P}\right) \qquad (19)$$

where $(T_0, P_0)$ are VFT singular temperature and pressures, located in the solid glass phase.

$D_T, D_P$ are fragility strength coefficients related to the temperature of pressure path of approaching the glass transition.

Despite the success of VFT equation in empirical applications and model analysis,[2-4,12,13] its general fundamental validity has been essentially questioned recently.[5,46,47] However, glycerol can be encountered to the limited group of materials where the VFT parameterization remains valid.[47] With the SA and SB behavior is inherently associated with the concept of fragility, one of the most prominent ideas within the glass transition physics.[2,4,12,13] The fragility constitutes the metrics of the 'degree' of the SA or SB behavior over basic Arrhenius or Barus ones. It is defined via the temperature related isobaric fragility (eq. (2)) and for the isothermic, pressure related path as: $m_{T=const} = \left[d\log_{10} x(P)/d(P/P_g)\right]_{P=P_g}$.[13,45,48,49]

In ref.[45] following links between $x(T, P)$ experimental data and basic parameters describing SA or SA dynamics were derived:

$$d\ln x(T)/d(1/T) = \Delta H_a^x/R = T m_p/\log_{10} e \quad \text{for } P = const \qquad (20)$$

and $d\ln x(P)/dP = \Delta V_a^x/RT = m_T/\log_{10} e$ for $T = const$. $\qquad (21)$

Their substitution into eq. (13) one obtains the link between fragility and the FDSE exponent:

$$S = \frac{\Delta H_a^\sigma}{\Delta H_a^\tau} = \frac{m_P^\sigma}{m_P^\tau} \quad (P = const) \quad \text{and} \quad S = \frac{\Delta V_a^\sigma}{\Delta V_a^\tau} = \frac{m_T^\sigma}{m_T^\tau} \quad (T = const) \qquad (22)$$

**Experimental**

Fluid nanocomposite/nacolloidal mixture with the concentration reaching 180 ppm of Ag (silver) nanoparticles in glycerol was prepared in the Institute of Superhard Materials in Kiev, Ukraine. It is notable that no additional chemicals or surfactants were needed to stabilize the nanocolloid and avoiding the sedimentation. Ag nanoparticles (AgNP) were synthesized via the localized ion-plasma sputtering and immediate implantation of freshly created nanoparticles to the carrier liquid in vacuum what allows to produce highly concentrated stable dispersions of ultra clean metals nanoparticles in various carrier liquids.[50] The size distribution of nanoparticle, averaged at ~ 25*nm* was below 2 %, what is shown on the Fig. 1. The concentration of Ag NP (180 ppm) was the highest for which long term stability (at least 1 year) was reached, without any additional component. However, a further increase of nanoparticles concentration led to their aggregation. Fig. 1 also contains two photos: (i) the bottle with the nanocolloid and (ii) the SEM picture (ambient conditions). For the latter, the view is influenced by the preparatory to SEM treatment and the fact that several layers of Ag NP are 'collected" on a plane.

The quality of nanoparticles was comparable with the ultraclean solutions produced by laser ablation in liquids (LAL) but the combined ion-plasma sputtering demonstrate much higher productivity and cost effectiveness.[50,51] Notwithstanding, immediately prior to measurements Ag nanocolloid samples were ultrasound sonicated for few hours, to preserve additionally uniform dispersion of nanoparticles.

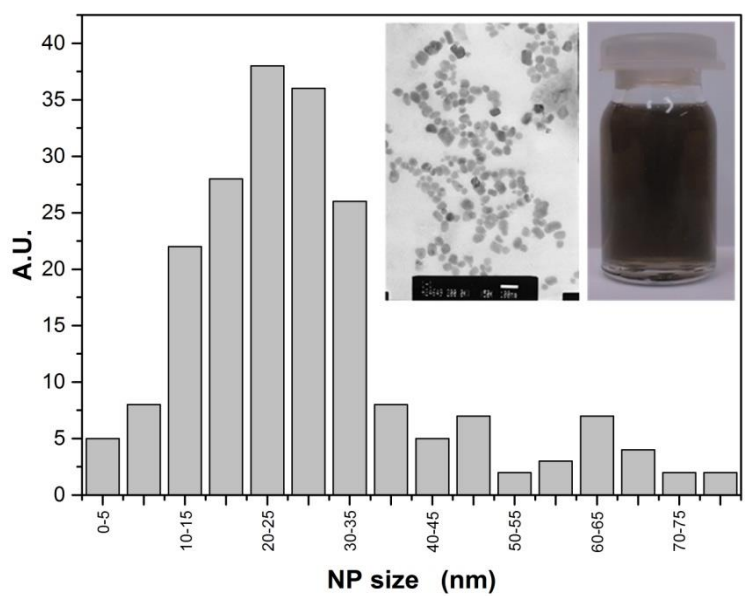

**Figure 1**. *The size distribution of Ag nanoparticles. Inset show the view of Ag dispersion in glycerol (left) and the SEM picture (right). The visible arrangement of nanoparticles can be considered as the result of sample preparations for SEM visualization.*

Dynamics of the pressurized glycerol and Ag-glycerol nanocolloid were tested via the piston-based high pressure set-up, described in ref.[52] The gap of the flat parallel measurement capacitor was equal to 0.2 mm. The macro-size of the gap made it possible to reduce parasitic artifacts associated with gas bubbles, finite dimensions or very large intensities of the measurement electric field, which appears for micrometric gaps.

The BDS spectrum, was monitored using the BDS Alpha Novocontrol spectrometer giving permanent 6 numbers resolution for imaginary and real part of dielectric permittivity. This report focuses on the pressure evolution of DC conductivity $\sigma$ and the primary relaxation time $\tau_\alpha$. The latter was estimated directly from peak frequency of dielectric loss curves via $\tau_\alpha = 1/2\pi f_{peak} = 1/\omega_{peak}$. The DC conductivity from the low frequency increase via the dependence $\varepsilon''(f) = \sigma/\varepsilon_0 \omega$.[2,12] Typical BDS spectra obtained and analyzed within the given research, characteristic both for temperature and pressure studies, are shown in Fig. 2.

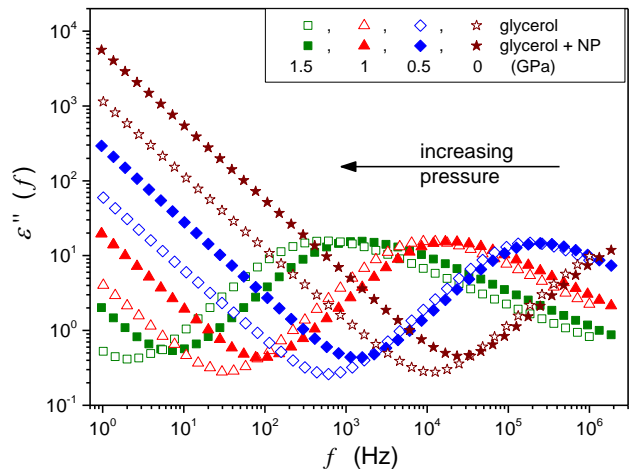

**Figure 2** *The behavior of the imaginary part of dielectric permittivity in glycerol and glycerol + silver nanoparticle nanocolloid. Notable is the increasing impact of Ag nanoparticles on compressing.*

It is worth stressing the BDS offers unique possibilities of high resolution studies both vs. temperature at atmospheric pressure and under high hydrostatic pressures. Moreover, the fact that both $\sigma(T,P)$ and $\tau(T,P)$ data can be determined from the same $\varepsilon''(f)$ scan notably reduces parasitic, biasing, artifacts. So far, BDS studies under high pressures are limited to frequencies $f < 1 \div 10 MHz$, due to still unsolved technical problems.[13] Nevertheless this frequency/time domain clearly correlates with the range of the ultraviscous/ultraslowing domain in glass forming systems. There is a broad evidence of BDS studies under atmospheric pressure focused on FDSE eqs. (3) and (9),[2,13,16-41] but surprisingly there are still no results for glycerol, which is one of canonical glass forming liquids. This can be associated with the fact that glycerol can be encountered as, so called, strong glass formers, for which the clear manifestation of the SB behavior needs studies well above the moderate range of pressures used so far.

This report presents results of the first ever FDSE focused test entering the multi GPa domain. High pressure BDS studies were carried out using the innovative piston-based method for $T \approx 258K$ isotherm, well below the room temperature tests dominated so far.[17-24] For the selected isotherm the 'glass pressure' can be estimated as $P_g = 1.95$ GPa, following $T_g(P_g)$ diagram presented in ref.[53]. The precision of pressure estimation was equal to 1 MPa and 0.02 K for temperature. All results were reversible (i.e. they could be obtained both on cooling and heating as well as compressing and decompressing). It is notable that earlier FDSE pressure studies were limited to very fragile (strongly SB) glass formers for which applied (moderate) pressures were able to induce significant changes of the time-scale.[2,13,18,10-26,30,39] For results presented below a similar time-scale was obtain also for glycerol, due the extension of the range of pressures up to challenge 1.5 GPa. There have been no reports regarding FDSE behavior in nancolloids/nacocomposites hitherto.

**Results and Discussion**

This report focuses on the FDSE behavior in the nanocolloid composed of glycerol and Ag nanoparticles (NP). The 'background" behavior in ultraviscous glycerol, which has been lacked so far, is also discussed.

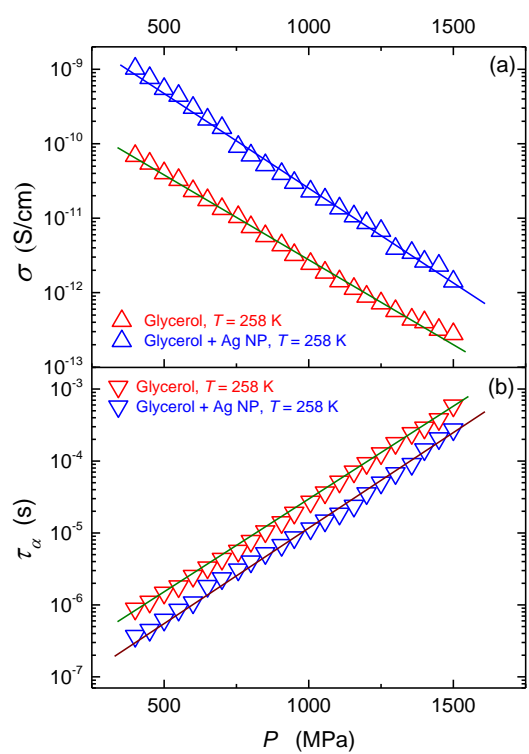

**Figure 3**  *The pressure evolution of the primary relaxation time (τ) and DC conductivity (σ) for pure glycerol and glycerol + Ag NP composite. Lines are guides for eyes.*

Fig. 3 shows pressure evolutions of primary relaxation times and DC electric conductivity in glycerol and its nanocolloid with Ag NP under compression, in agreement with Fig. 2. The addition of nanoparticles notably increases electric conductivity (Fig. 3a, ~ decade) and decreases the primary relaxation time (Fig. 3b, ~ half of decade). The same pattern takes place for the temperature behavior under atmospheric pressure. The The large enhancement in the thermal conductivity and electric conductivity when even a small amount of metallic and other nanoparticles is dispersed is well evidence for lots of systems.[54-58] These extraordinary features have led to a set of innovative applications and to the emergence a fluid created a large area of practical application and a new area of research recalled as '*nanofluidics*".[59] Several efforts have been made to explain conductivity enhancements in fluids due to the addition of nanoparticles. However, there has been no general consensus on this issue despite their practical significance.[56-59] No ultimate theoretical model is also available to predict nanofluid viscosity with good accuracy.[58,59] Generally, the addition of nanoparticles increase the viscosity of the resulted nanofluid, what is linked to nanoparticles aggregation.[59] However, in viscous heavy oils adding of nanoparticles can notably reduce the viscosity.[60-62] This unique behavior is indicated as particularly important for petroleum industry.[60] Following eqs. (4), (6) the same pattern may be expected for viscosity and primary relaxation time. For the latter, the direct experimental evidence is very poor. Notwithstanding., for few 'dense" fluid systems the atypical increase of electric conductivity matched with the decrease of the primary relaxation time have been reported.[61-64] This report presents the first ever results for an ultraviscous glass forming nanocolloid/nanofluid, supercooled and superpressed.

Generally, in very viscous and ultraviscous systems the self- aggregation of nanoparticles can be difficult and the sonication can further support the stable and homogeneous dispersion of nanoparticles. Consequently, the self-aggregation most often observed in 'typical' nanofluids can be limited or even avoided. The behavior of the ultraviscous liquids near the glass transition is dominated by the emergence of 'dynamic heterogeneities' with larger density than the fluid-like surrounding and even possible elements of structural arrangements.[2] This can cause the collection of nanoparticles on the border of solid-like heterogeneities and the fluid-like surrounding and subsequently the fragmentation of 'heterogeneities'.

Consequently, smaller number of molecules is located within heterogeneities what can lead to the decrease of the average rotational relaxation time. This can support also the decrease of viscosity. The hypothetical string-like arrangements of dynamic heterogeneities in ultraviscous domain near the glass transition[2] can facilitate string like arrangements of nanoparticles. Such behavior can support larger electric and heat conductivity. The decrease of viscosity can be also supported by the appearance of string-like, elongated mesoscale structures in a way similar as the addition of a selected polymer to a fluid. It is notable that the increase of electric conductivity matched with the decrease of the relaxation time and viscosity was observed in nanoparticles doped liquid crytals,[63,64] in which the behavior is dominated by multimolecular, pretransitional fluctuations.[65]

Figure 4 shows interplay between the translational and orientaional dynamics in pure and Ag NP doped glycerol. The analysis shows that in glycerol for $T_B < T < T_g$ the FDSE exponent $S \approx 1$. Hence, in glycerol solely a negligible decoupling between the DC conductivity and the relaxation time takes place. This behavior is atypical, taking into account the dominating evidence for other glass forming liquids, indicating that $S < 1$ for the utraviscous region.

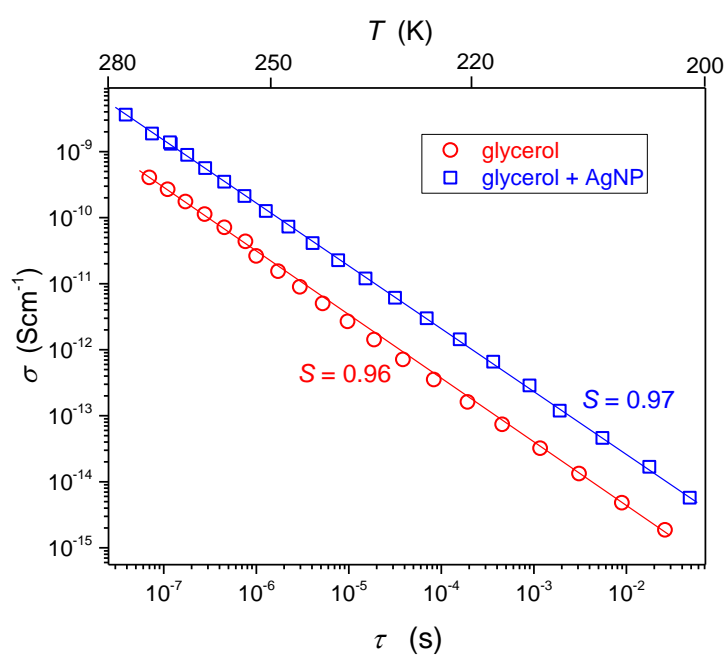

**Figure 4** *Temperature test of DSE law in pure glycerol and glycerol + AgNP composite at the pressure P = 0.1 MPa. The right scale is for the nanocolloid and the left one for glycerol. Slopes of lines, determining FDSE exponent, are also given*

Results of the analysis of the translational-orientational decoupling on compressing up to $P = 1.5$ GPa is presented in Fig. 5 for glycerol and in Fig. 6 for the nanocolloid. For glycerol up to $P \sim 1 GPa$ the FDSE exponent $S \approx 1$, i.e. the behavior is resembles one observed under atmospheric pressure. However, on further pressurization towards the glass transition the gradual translational – orientational decoupling towards the exponent $S \approx 0.75$ occurs. For glycerol plus Ag NP nanocolloid this transformation is 'sharp" and occurs at well-defined pressure $P \approx 1.2 GPa$, where a jump from $S \approx 1$ to very extremely decoupled FDSE behavior with $S \approx 0.5$ takes place.

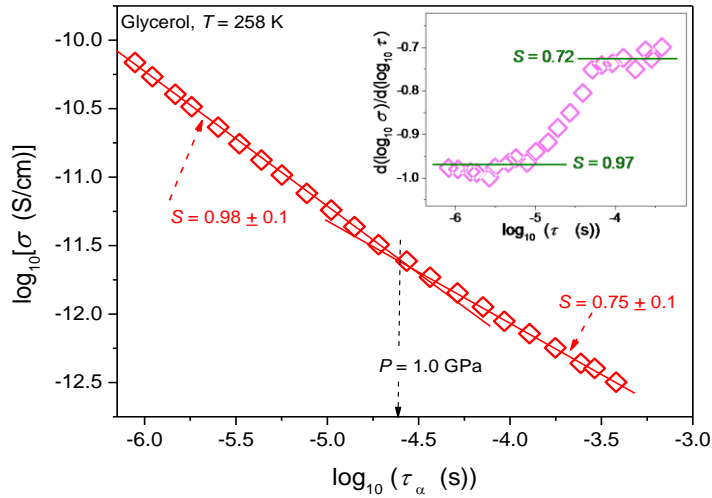

**Figure 5** *Test of the fractional Debye-Stokes-Einstein behavior in pressurized glycerol at T=258 K. The inset shows results of the derivative-based, distortions sensitive analysis of data from the main part of the plot.*

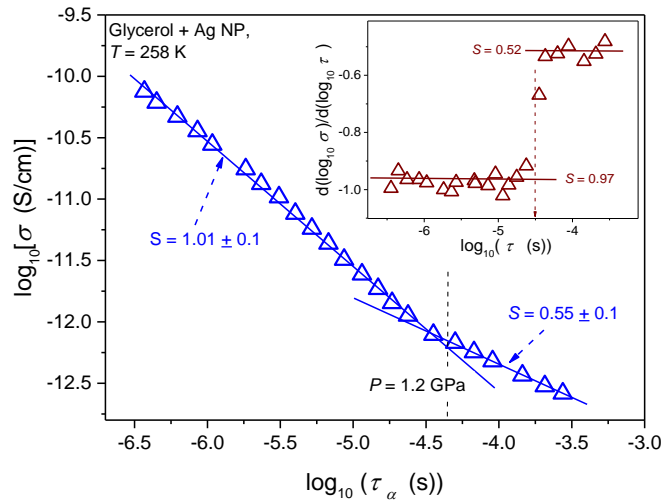

**Figure 6** *Test of the fractional Debye-Stokes-Einstein behavior in pressurized nanocolloid (glycerol + AgNP) at T=258 K. The inset shows results of the derivative-based, distortions sensitive analysis of data from the main part of the plot.*

The FDSE behavior observed under pressure can be correlated with the broadening of primary relaxation loss curves, as shown in Fig. 7. Up to $P \approx 1 GPa$ the clear superposition of loss curves takes place. For higher pressures the broadening, both for the high- and low-frequency branches of loss curves, occurs.

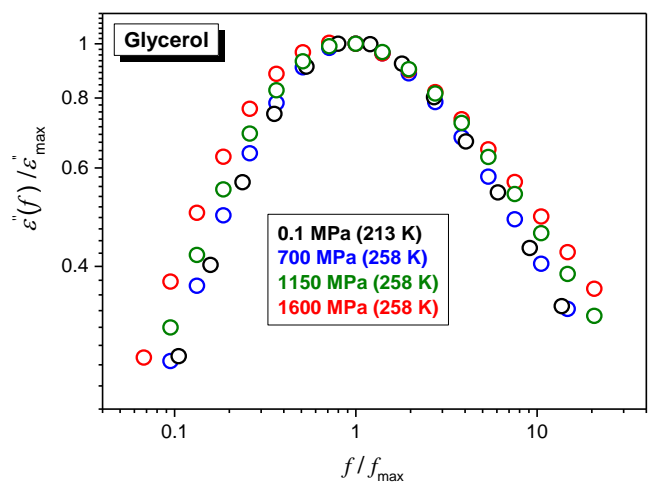

**Fig. 7** *Normalized superposition of dielectric loss curves $\varepsilon''(f)$ for ultraviscous glycerol under various pressures.*

The comparison of $\varepsilon''(f)$ evolution in Fig.2 and Fig.7 enables a qualitative explanation of FDSE coupling/decoupling manifesting via eqs. (4) and (13). In Fig. 1 the primary relaxation time is determined from coordinates of the loss curve peak as $\tau = 1/\omega_m = 1/2\pi f_{peak}$ and the DC electric conductivity is using the plot: $y = \log_{10} \varepsilon''(\log_{10} f) = ax + b = a(\log_{10} f) + b$ with the slope $a = 1$, from the linear behavior in the low frequency part of the spectrum. If the low frequency branch of $\varepsilon''(f)$ loss curves do not change the shift of $\sigma(T,P)$ exactly follows the shift of $\tau(T,P)$ on cooling or pressuring. However, the broadening of $\varepsilon''(f)$ on cooling or pressuring loss curve induces an extra shift of $\sigma(f)$, what can result in FDSE decoupling via eq. (13).

**Conclusions**

Glycerol is a versatile material due to its enormous significance in a variety of applications ranging from biotechnology to pharmacy, cosmetics, "*green and biodegradable*" plastics, textiles and foodstuffs industries.[66-69] Glycerol and Ag nanoparticles based nanocolloids/nanocomposites may appear important in these applications due to well known great antimicrobial activity of Ag nanoparticles.[70,71]

From the fundamental point of view glycerol has a simple molecular structure, large permanent dipole moment and the relatively small electric conductivity, what coincides with preferred features for the broad band dielectric spectroscopy (BDS) monitoring.[12] It can be also very easily supercooled. All these caused that glycerol has gained the position of a model "classical" system in glass transition studies.[2,12,13] This report present also the first ever experimental report of FDSE behavior in an nanocolloid/nanocomposite system. Also the range of implemented high pressure is well above earlier studies. The key results of the report is the crossover from the almost coupled (DSE) to the strongly decoupled (FDSE) behavior for the pressurized glycerol and glycerol based nanocolloid in the vicinity of the glass transition. For the extremely decoupled state associated with $S \approx 0.5$, which is probably the lowest value of the FDSE exponent detected so far. The crossover takes place within the nonergodic ultraviscous domain where so far it was observed solely for the transition into the ultraviscous domain (ergodic – nonergodic transformation) at much larger distances from the glass transition. The crossover within the pressurized ultraviscous domain takes place both for the 'pure' pure glycerol ($S \approx 1 \rightarrow S \approx 0.7$) and glycerol + AgNP nanocolloid ($S \approx 1 \rightarrow S \approx 0.5$). For the latter it took place for one, well defined pressure. Hence, the presence of nanoparticles leads to a qualitative enhancement of manifestations of this phenomenon. The explanation of origin of this phenomenon needs further experimental studies able to follow nanoparticles in the immediate vicinity of the glass transition under high pressure, what in fact is beyond the current experimental state-of-the-art. New possibilities can open advanced microscopic observations of highly compressed liquids, based on the set-up currently build in the lab of the authors. One of speculative explanations can be related to better definition of heterogeneities due to the inclusion of nanoparticles. Their possible chain-like arrangements can create elements of uniaxial, orientational ordering within heterogeneities. As shown recently the fragility is proportional to the parameter $n$ describing the local, symmetry $m \propto n$, with $0.2 < n < 1.5$: the lower value is for the dominated positional ordering, the upper limit is for the clearly orientational case and $n = 1$ is for the 'no-symmetry" case. It is also notable that pressurization notably increases density (for glycerol in the GPa domain down up to 20 %), thus decreasing inter-particle distances what can facilitate ordering of Ag nanoparticles. Worth recalling is also Fig. 7 and discussion nearby indicating the link of the decrease of the FDSE exponent near the glass temperature to the broadening of the distribution of primary relaxation time, which is strongly linked to the enhancement of appearance of dynamic heterogeneities.

It is noteworthy that the analysis based on $D(T)$, $\eta(T)$ and experimental data indicated for glycerol FDSE exponent $\zeta \approx 0.85$ (eqn 4),[8] the value suggested as universal ones for the ultraviscous domain. This report does not confirm this finding. It has been found that glycerol exhibit a unique behavior: (i) first, the is no change of FDSE exponent when passing the dynamic crossover point, namely $S \approx 1$ both below and above $(T_B, P_B)$, (ii) a new (not observed so far) crossover to the behavior governed by $S < 1$ occurs already within the ultaviscous domain, particularly under high compression and (iii) the presence of Ag nanoparticle in glycerol notably strengthen features related to the FDSE domain emerging in the immediate vicinity of the glass transition.

The simple analysis based solely on SA and SB relations and the general FDSE dependences yielded a new link between the FDSE exponent $S$ and temperature and pressure related fragilities indicating that for the $m_T^\sigma = S m_T^\tau$ and $m_P^\sigma = S m_P^\tau$. Larger fragility coefficient for primary relaxation time related processes is in agreement with discussed above fundamental findings of ref. [43]. Worth stressing is the obtained link of the FDSE exponent to the apparent activation enthalpy and volume.

Concluding, this report shows new features of translational-orientational decoupling dynamics emerging from the impact of very high pressures on ultraviscous glycerol and the formation of Ag NP based nanocolloid. The report also indicates that some basic feature of the decoupling can be deduced from the general super-Arrhenius, super-Barus and FDSE equations.

**Acknowledgements**

SJR research was carried due to the support of the National Science Centre (Poland) via grant 2011/01/B/NZ9/02537. ADR and SSz were supported by the National Centre for Science (Poland), grant UMO. 2011/03/B/ST3/02352.

**References**

1. D. Kennedy, *Science*, 2005, **309**, 83.

2. K. L. Ngai, *Relaxation and Diffusion in Complex Systems*, Springer, Berlin, 2011.

3. R. A. L. Jones, *Soft Condensed Matter Physics*, Oxford Univ. Press., Oxford, 2002.

4. C. A. Angell, *Strong and fragile liquids, in: Relaxations in Complex Systems*, eds.: K. L. Ngai and G. B. Wright, National Technical Information Service, U.S. Dept. of Commerce, Springfield,1985.

5. J. C. Martinez-Garcia, S. J. Rzoska, A. Drozd-Rzoska, S. Starzonek, and J. C. Mauro, 2015, *Sci. Rep.* **5**, 8314 (1-7).

6. V. Agapov, A. L., Novikov, V. N. & Sokolov, A. P. *Fragility and other properties of glass-forming liquids: Two decades of puzzling correlations*, in L. A. Greer, K. Kelton, S. Sastry (eds.*), Fragility of glass forming liquids*, Hindustan Book Agency, 2013.

7. A. I. Nielsen, S. Pawlus, M. Paluch, and J. C. Dyre, 2008, *Philos. Mag.* **88**, 4101-4108.

8. F. Mallamace, C. Branca, C. Corsaro, N. Leone, J. N. Spooren, S.-H. Chen and H. E. Stanley, *Proc. Natl. Acad. Sci.*, 2010, **28**, 22457-22462.

9. M. A. Anisimov, *Critical Phenomena in Liquids and Liquid crystals,* Gordon and Breach, Reading, 1992.

10. V. N. Novikov and A. P. Sokolov, *Phys. Rev. E*, 2003, **67**, 031507.

11. P. J. Griffin J. R. Sangoro, Y. Wang, A. P. Holt, V. N. Novikov, A. P. Sokolov, M. Paluch and F. Kremer, *Soft Matter*, 2013, **43**, 10373-10380.

12. F. Kremer and A. Schoenhals, *Broadband Dielectric Spectroscopy*, Springer Verlag, Berlin, 2003.

13. G. Floudas, M. Paluch, A. Grzybowski, K. L. Ngai, *Molecular Dynamics of Glass-Forming Systems: Effects of Pressure*, Springer, Berlin, 2012.

14. K. A. Dill and S. Bromber, *Molecular Driving Forces: Statistical Thermodynamics in Chemistry and Biology*, Garland Science, London, 2011.

15. J. C. Dyre, *Rev. Mod. Phys.*, 2006, **78**, 953-972.

16. L. Andreozzi, A. Di Schino, M. Giordano and D. Leporini, *Europhys. Lett.*, 1997, **38**, 669-674.

17. S. Corezzi, M. Lucchesi, P. A. Rolla, *Phil. Mag. B*, 1999, **79**, 1953-1963.

18. S. Hensel Bielowka, T. Psurek, J. Ziolo, and M. Paluch, *Phys. Rev. E*, 2001, **63**, 062301.

19. P. G. Debenedetti, T. M. Truskett, and C. P. Lewis, F. H. Stillinger, *Theory of supercooled liquids and glasses. Energy Landscape and Statistical Geometry Perspective,* pp.21-30, in *Advances in Chemical Engineering,* vol. (28), Academy Press., Princeton, 2001.

20. M. Paluch, T. Psurek, and C. M. Roland, *J. Phys.: Condens. Matt.*, 2002, **14**, 9489-9494.

21. M. Paluch, C. M. Roland, J. Gapinski and A. Patkowski, *J. Chem. Phys.*, 2003, **118**, 3177-3186.

22. R. Casalini and C. M. Roland, *J. Chem. Phys.*, 2003, **119**, 4052-4059.

23. R. Casalini, M. Paluch, T. Psurek, and C. M. Roland, *J. Mol. Liq.*, 2004, **111**, 53-60.

24. T. Psurek, J. Ziolo, M. Paluch, *Physica A*, 2004, **331**, 353-364.

25. G. P. Johari and O. Andersson, *J. Chem. Phys.*, 2006, **125**, 124501(7).

26. O. Andersson, G.P. Johari, R.M. Shanker, *J. Pharm. Sci.*, 2006, **95**, 2406-2418.

27. R. Richertand, K. Samwer, *New J. Phys.*, 2007, **9**, 36(11).

28. G. Power, J. K. Vij, and G. P. Johari, *J. Phys. Chem. B*, 2007, **111**, 11201-11208.

29. M. G. Mazza, N. Giovambattistam H. E. Stanley, F. W. Starr, *Phys. Rev. E*, 2007, **76**, 031203.

30. S. Pawlus, M. Paluch, J. Ziolo, C. M. Roland, J. *Phys.: Condens. Matt.*, 2009, **21**, 332101.

31. L. Xu, F. Mallamace, Z. Yan, F. W. Starr, S. V. Buldyrev and H. E. Stanley, *Nature Physics*, 2009, **5**, 565-569.

32. A. Drozd-Rzoska and S. J. Rzoska, *Anomalous Decoupling of the dc Conductivity and the Structural Relaxation Time in the Isotropic Phase of a Rod-Like Liquid Crystalline Compound*, pp. 141-152, in *Metastable Systems under Pressure*, ed. by S. J. Rzoska, V. Mazur and A. Drozd-Rzoska, Springer, Berlin, 2010.

33. G. Sesé, J. O. de Urbina, and R. Palomar, *J. Chem. Phys.,* 2012, **137**, 114502 .

34. K. V. Edmonda, M. T. Elsesserb, G. L. Huntera, D. J. Pinebo, and E. R. Weeksa, *Proc. Natl. Acad. Sci.*, 2012, **109**, 17891-17896.

35. F. Mallamace, C. Corsaro, N. Leone, V. Villari, N. Micali and S.-H. Chen, *Sci. Rep.*, 2014, **4**, 3747.

36. Z. Shi, P. G. Debenedetti, and F. H. Stillinger, *J. Chem. Phys.*, 2013, **138**, 12526.

37. N. Leone, V. Villari, N. Micali&S.-H. Chen, *Sci. Rep.*, 2013, **4**, 3747(1-8).

38. J. Swiergiel and J. Jadzyn, *Phys. Chem. Chem. Phys.*, 2011, **13**, 3911–3916

39. Z. Wojnarowska, Y. Wang, J. Pionteck, K. Grzybowska, A. P. Sokolov, and M. Paluch, *Phys. Rev. Lett.*, 2013, **111**, 225703.

40. E. Herold, M. Strauch, D. Michalik, A. Appelhagenand, R. Ludwig. *Phys. Chem. Chem. Phys.*, 2014, **15**, 3040-3048.

41. D. A. Turton, K. Wynne, *J. Phys. Chem. B*, 2014, **118**, 4600–4605.

42. S. J. Rzoska and V. Mazur (eds.), *Soft Matter under Exogenic Impacts*, Springer, Berlin, 2007.

43. A. P. Sokolov, K. S. Schweitzer, *Phys. Rev. Lett.,* 2009, **102**, 23831.

44. C. Barus, *Proc, Am. Acad.*, 1891, **27**, 13-15.

45. A. Drozd-Rzoska and S. J. Rzoska*, Phys. Rev. E*, 2006, **73**, 041502.

46. J. C. Martinez-Garcia, S. J. Rzoska, A. Drozd-Rzoska, J. Martinez-Garcia, *Nat. Commun.*, 2013, **4**, 1823 -1823.

47. J. C. Martinez-Garcia, S. J. Rzoska, A. Drozd-Rzoska, J. Martinez-Garcia, and J. C. Mauro, *Sci. Rep*. 2014, **4**, 5160 (1-7).

48. M. Paluch, S. J. Rzoska, J. Zioło, *J. Phys.: Condens. Matt.*, 1998, **10**, 4131-4138.

49. A. Drozd-Rzoska, S. J. Rzoska and C. M. Roland, *J. Phys.: Condens. Matt.*, 2008, **20**, 244103-244111.

50. L. Barber, T. Kistersky, Patent 5591313 USA, IC C 23 C 14/04. *Apparatus and method for localized ion sputtering*/ Publ. 7.01.97.

51. P. Liu, Y. Liang, H.B. Li, G.W. Yang, *Nanomaterials: Applications and Properties,* vol. 1, pp. 136 -150, NAP, Beijing, 2011.

52. M. Paluch, M. Sekula, S. Pawlus, S. J. Rzoska, J. Ziolo, C. M. Roland, *Phys. Rev. Lett.* 2003, **90**, 175702.

53. A. Drozd-Rzoska, S. J. Rzoska, M. Paluch, A. R. Imre, C. M. Roland, *J. Chem. Phys.* 2007, **126**, 164504 (2007).

54. S. U. S. Choi, Z. G. Zhang, W. Yu, F. E. Lockwood, E. A. Grulke, *Appl. Phys. Lett.* 2001, **79**, 2252-2254.

55. J. A. Eastman, SUS Choi, S Li, W. Yu, W. Thompson: *Appl. Phys. Lett.* 2001, **78**, 718

56. P. Warrier and A. Teja, *Nanoscale Res. Lett.* 2011, **6**, 247-253.

57. I. M. Mahbubul, R. Saidur, M. A. Amalina, *Int. J. Heat and Mass Transfer* 2012, **55**, 874–888.

58. S. B. White, A. Jau-Min Shih and K. P. Pipe, *Nanoscale Res Lett.* 2011, **6**, 346 (1-5)

59. S. K. Das, S. U. Choi, W. Yu, T. Pradeep, *Nanofluids: Science and Technology*, Wiley, NY, 2008.

60. Y. H. Shokrlu, T. Babadagli, *J. Petrol. Sci. And Engn.* 2014, **119**, 210-220.

61. R. J. Sengwa, S. Choudhary, S. Sankhla, *eXPRESS Polym. Lett.* 2008, **2**, 800–809.

62. M. Dong, L. P. Shen, H. Wang, H. B. Wang, and J. Miao, *J. Nanomat.* 2013, **2013**, 1-7.

63. S. P. Yadav, M. Pande, R. Manohan, S. Singh, *J. Mol. Liq.* 2015, **208**. 34-37.

64. S. Krishna Prasad, M. Vijay Kumar, T. Shilpa and C. V. Yelamaggad, *RSC Adv.* 2014, **4**, 4453–4462.

65. S. J. Rzoska, A. Drozd-Rzoska, P. K. Mukherjee, D. O. Lopez, J. C. Martinez-Garcia, *J. Phys.: Condens. Matt.* 2013, **25**, 245105 (11).

66. M. Pagliaro and M. Rossi, *Future of Glycerol: New Usages for a Versatile Raw Material*, RSC, Cambridge, 2008.

67. E. S. Stevens, *Green Plastics: An Introduction to the New Science of Biodegradable Plastics*, Princeton Univ. Press, Princeton, 2002.

68. F. Gunstone, *Oils and Fats in the Food Industry*, J. Wiley & Sons, NY, 2009.

69. J. Choi, J. C. Bishof, *Cryobiology* 2010, **60**, 49-52.

70. C. Balny and R. Hayashi, *High Pressure Bioscience and Biotechnology*, in series *Progress in Biotechnology* vol. 13 (Elsevier, Amsterdam, 1996).

71. J. A. Lemire, J. J. Harrison, R. J. Turner, *Nat. Rev. Microbiol.,* 2013, **11**, 371−384.